# Non-monotonic Evolution of the Blocking Temperature in Dispersions of Superparamagnetic Nanoparticles

D. Serantes, D. Baldomir, and M. Pereiro

*Instituto de Investigacións Tecnolóxicas and Departamento de Física Aplicada, Universidade de Santiago de Compostela, 15782 Santiago de Compostela, Galiza (Spain)*

C. E. Hoppe

*Div.Polímeros, INTEMA (UNMdP-CONICET) Universidad Nacional de Mar del Plata B7608FDQ, Mar del Plata, Argentina*

F. Rivadulla

*Departamento de Química Física, Universidade de Santiago de Compostela, 15782 Santiago de Compostela, Spain*

J. Rivas

*Departamento de Física Aplicada, Universidade de Santiago de Compostela, 15782 Santiago de Compostela, Spain*

*Abstract*

We use a Monte Carlo approach to simulate the influence of the dipolar interaction on assemblies of monodisperse superparamagnetic $\gamma$-$Fe_2O_3$ nanoparticles. We have identified a critical concentration $c^*$, that marks the transition between two different regimes in the evolution of the blocking temperature ($T_B$) with interparticle interactions. At low concentrations ($c < c^*$) magnetic particles behave as an ideal non-interacting system with a constant $T_B$. At concentrations $c > c^*$ the dipolar energy enhances the anisotropic energy barrier and $T_B$ increases with increasing $c$, so that a larger temperature is required to reach the superparamagnetic state. The fitting of our results with classical particle models and experiments supports the existence of two differentiated regimes. Our data could help to understand apparently contradictory results from the literature.

PACS: 75.75.+a, 75.20.-g, 75.40.Mg, 75.50.Tt

## I. INTRODUCTION

Envisaging new methods for the synthesis of monodisperse magnetic nanoparticles (NPs) constitutes one of the most active and challenging research fields in materials science due to the almost infinite uses foreseen for these systems in biomedicine,[1] magnetic recording,[2] energy production, etc.[3] Many of these applications rely on the possibility of obtaining well dispersed assemblies of NPs in a non-magnetic matrix (normally diamagnetic solids like polymers, $SiO_2$, etc). This step can be really tough due to the high tendency of NPs to aggregate producing heterogeneous and uncontrolled structures. Aggregation is typically induced by interaction forces (van der Waals, dipolar) that depend on different variables of the system, like the nature of the matrix and the molecules that coat the surface of the particles. In this sense, determining the magnetic properties of these dispersions of NPs, particularly the magnetization under *zero-field-cooling* and *field-cooling* conditions (ZFC-FC) and the blocking temperature ($T_B$), is becoming an increasingly popular analytical technique, due to its high sensitivity to the distribution of the particles in the matrix, its surface oxidation state, homogeneity, etc. However, it is usual to find in the literature dissimilar (sometimes contradictory) results from, a priori, similar samples. This is most probably due to the high sensitivity of the magnetic properties to interparticle interaction, which can make very easy to confuse extrinsic effects with intrinsic ones. In fact, understanding the role played by dipolar interactions into the magnetic behavior of the system remains a challenge despite the intense investigation and discussion devoted to it.[4,5,6] The interest of solving it is important for practical applications since it is a key-parameter for driving the magnetic response of nanotechnological devices and applications using magnetic NPs.

Here we describe the effect of the magnetic dipolar interaction in the evolution of the ZFC-FC curves and the $T_B$ of a monodisperse system of maghemite-like superparamagnetic (SPM) NPs. The anisotropy energy (responsible for the existence of $T_B$) is for these particles rather low, and so the system is very sensitive to variations of the dipolar energy. We will compare our results from Monte Carlo (MC) simulations with our own experiments and other ones available in the literature. Our results demonstrate the existence of two different regimes in the evolution of $T_B$ with the concentration of NPs in homogeneous dispersions.

## II. COMPUTATIONAL DETAILS

We have used a Monte Carlo (MC) technique[7,8,9] to study the magnetic response of an assembly of single domain magnetic nanoparticles as a function of the magnetic dipolar interactions between the particles. We have chosen for the system ideal characteristics accounting to eliminate deviations from the intrinsic behaviour, with the purpose to set an appropriate frame to understand the basic magnetic properties of the system. First, the particles are spatially distributed into an ideal liquid-like structure that resembles a magnetic ferrofluid without aggregations. To acquire such a distribution we have used a Lennard-Jones pair potential with periodic boundary conditions, in the same way as done in Ref. 7. The positions of the particles are kept fixed from now on, assuming the same condition of particles fixed in a non magnetic matrix or a frozen ferrofluid. The second idealization we make is to consider a wholly monodisperse system: the particles, characterized by its volume, anisotropy and magnetization, are assumed to be all equal in their characteristics, so that no effects on the magnetic response of the system may be attributed to polydispersity of any type.

The energies that we have taken into account to govern the magnetic behaviour of the system are anisotropy ($E_A$), Zeeman ($E_Z$), and dipolar ($E_D$). The anisotropy of the particles is assumed uniaxial for the sake of simplicity, so that for the *i*-particle $E_A^{(i)} = -KV\left(\vec{\mu}_i \cdot \hat{n}_i / |\vec{\mu}_i|\right)^2$, where K is the uniaxial anisotropy constant and $\hat{n}_i$ indicates the direction of the anisotropy easy axis. In the ideal superparamagnetic frame it is considered single-domain NPs with the inner atomic moments rigidly coupled, what results the total magnetic moment for the *i*-particle to be $|\vec{\mu}_i| = M_S V$, where $M_S$ is the saturation magnetization and V is the volume of the particle. The Zeeman energy is treated in the usual way $E_H^{(i)} = -\vec{\mu}_i \cdot \vec{H}$, and the dipolar interaction energy between two particles located at $\vec{r}_i$, $\vec{r}_j$ respectively, is given by $E_D^{(i,j)} = \left(\vec{\mu}_i \cdot \vec{\mu}_j\right)/r_{ij}^3 - 3\left(\vec{\mu}_i \cdot \vec{r}_{ij}\right)\left(\vec{\mu}_j \cdot \vec{r}_{ij}\right)/r_{ij}^5$, with $\vec{r}_{ij}$ the vector connecting the particles. To evaluate the long-range interactions we applied periodic boundary conditions by means of Ewald's summation. The total energy of the system is the summation of the different terms extended to all the particles.

The motion of the individual magnetic moments of the particles as a function of the temperature (T) was driven by means of the Metropolis algorithm: in every MC step, we select a particle *i* at random and generate a new orientation of its magnetic moment. Then, we accept the new orientation with probability

min[1,exp(-ΔE/$k_B$T)], where ΔE is the energy difference between the attempted and present orientations, and $k_B$ is the Boltzmann constant. In every MC step N attempts are made, where N is the number of particles used in the simulation (N=125 in this study). The results are obtained averaging over 1000 different configurations, and extend those reported by us in Ref. 9 with a larger precision, in order to obtain more reliable conclusions.

We have studied the influence of the magnetic dipolar interaction on the superparamagnetic properties of the system by analyzing its influence on its characteristic blocking temperature ($T_B$), roughly evaluated as the temperature at which the ZFC curve exhibits a maximum. The strength of the dipolar interaction is introduced as proportional to the sample concentration, *c*. For a monodisperse particle assembly the equation relating the dipolar energy ($E_D$) and *c* is given by

$$E_D^{(i,j)} = g\sum_i \sum_{j\neq i}\left( \frac{\hat{e}_{\mu_i}\cdot \hat{e}_{\mu_j}}{a_{ij}^3} - 3\frac{\left(\hat{e}_{\mu_i}\cdot \vec{a}_{ij}\right)\left(\hat{e}_{\mu_j}\cdot \vec{a}_{ij}\right)}{a_{ij}^5}\right) \tag{5}$$

where $g = \frac{c}{c_0}\frac{KV}{N}$ characterizes the strength of the dipolar interaction. The dimensionless sample concentration *c* is the ratio between the total volume $\sum_i V_i$ occupied by the particles (NV for the monodisperse sample) and the volume $L^3$ of the sample, $c \equiv NV/L^3$ . The value $c_0 = \frac{2K}{M_S^2}$ is a dimensionless constant that characterizes the material.[8] The unit vector $\hat{e}_{\mu_i}$ stands for the direction of the magnetic moment $\mu$ of the particle *i*, and the reduced distance $\vec{a}_{ij}$ is defined as $\vec{a}_{ij} = \vec{r}_{ij}/L$ , the distance between the particles *i* and *j* divided by the size of the cubic box that contains the sample. The results obtained from our simulations are presented in reduced units directly related to the real ones, i.e., the reduced sample concentration is $c/c_0$, and $t=k_BT/2KV$ is the reduced temperature. The reduced applied magnetic field is $h=H/H_A$, where $H_A=2K/M_S$ is the anisotropy field of the particles, and the reduced magnetization is defined as $m = M/M_S \equiv \sum_i \cos\theta_i / N$ , with $\theta_i$ the angle between the magnetic moment of the *i*-particle and the direction of the applied magnetic field.

We studied the influence of the dipolar interaction on $T_B$ by simulating ZFC processes at different sample concentrations, ranging from the non-interacting diluted limit ($c/c_0 = 0.000$) to very dense samples ($c/c_0 \leq 0.320$). For every simulation, we have first demagnetized the samples at very high temperature and then cooled them down in zero applied magnetic field; once the sample has reached a very low temperature, a small reduced field h = 0.1 was applied and the magnetization was measured while the samples were heated up at constant rate until well above the reduced blocking temperature, $t_B$. The heating/cooling rate was $\Delta t=0.001225KV/k_B$ every 200 MC steps. The reduced susceptibility is defined as $\chi=m/h$. We systematically vary *c* to evaluate $t_B$ as a function of the dipolar interaction energy, as it is summarized in Fig. 2.[10]

## III. RESULTS AND DISCUSION

Some representative simulated ZFC susceptibility curves of the dispersions of magnetic nanoparticles at different sample concentrations are shown in Fig. 1, where also the FC curves of two different interacting conditions ($c/c_0 \rightarrow 0.000$ and $c/c_0 = 0.112$) are included to illustrate the reliability of the code. The applied magnetic field was $H = 0.1H_A$. This small value of the magnetic field was selected to not disturbing the intrinsic SPM behavior of the NPs. Large fields could mask this effect.

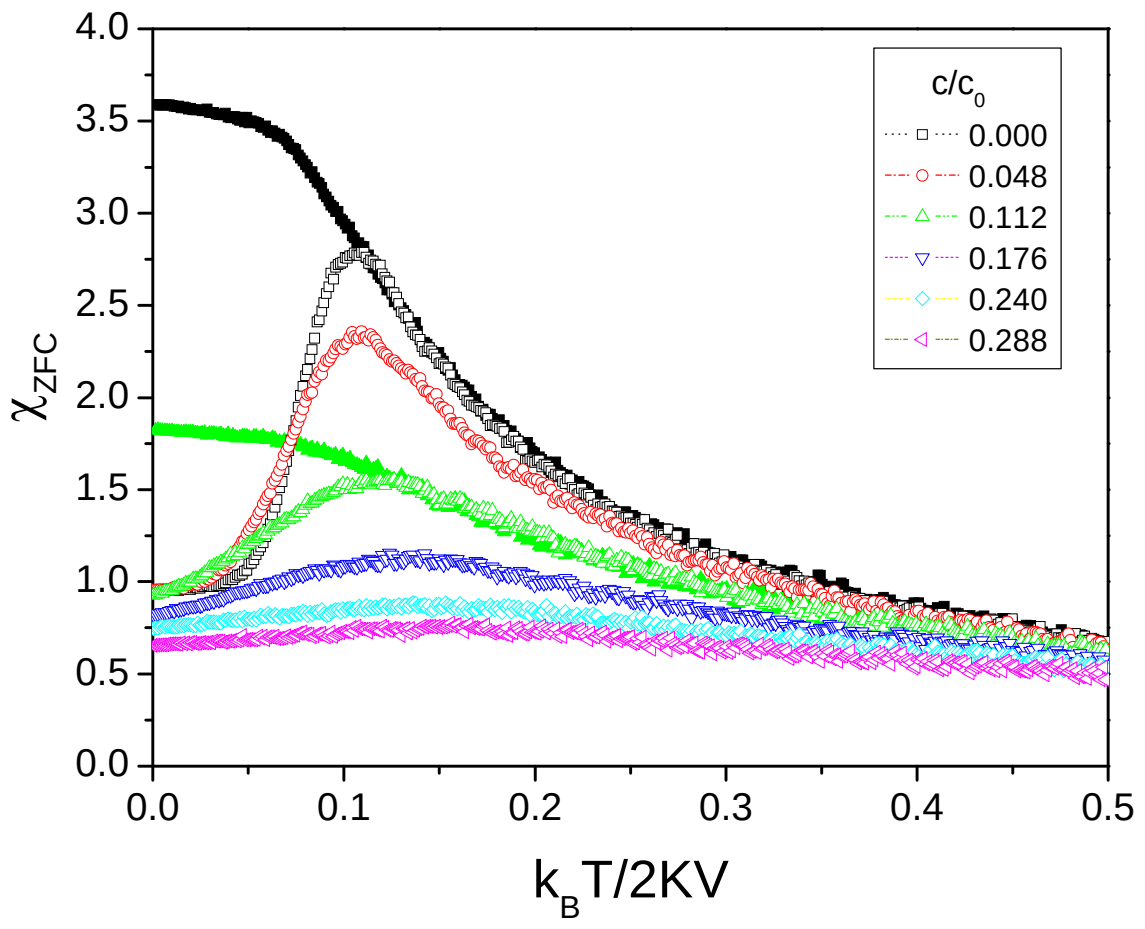

FIG. 1. ZFC curves (empty symbols) of some representative samples. Two FC curves (full symbols) are shown, corresponding to the non-interacting case ($c/c_0 \rightarrow 0.000$) and to $c/c_0 = 0.112$.

Numerical simulations of the ZFC/FC processes show a good agreement with the general trend of experimental results on SPM particles.[11,12] The ZFC curves exhibit a maximum at the reduced blocking temperature ($t_B$) that marks the transition to the SPM regime. Above $t_B$ it is observed the superposition of the FC with the ZFC curves that shows the reversible character of the SPM behavior. Below $t_B$, in the irreversible range, the FC curve separates from the ZFC. We also see from Fig. 1 an increase of $t_B$ as the concentration increases, in agreement with the results reported in experimental[12] and theoretical works.[7] The detailed dependence of $t_B$ vs. $c/c_0$ extracted from a complete set of measurements is summarized in Fig. 2.

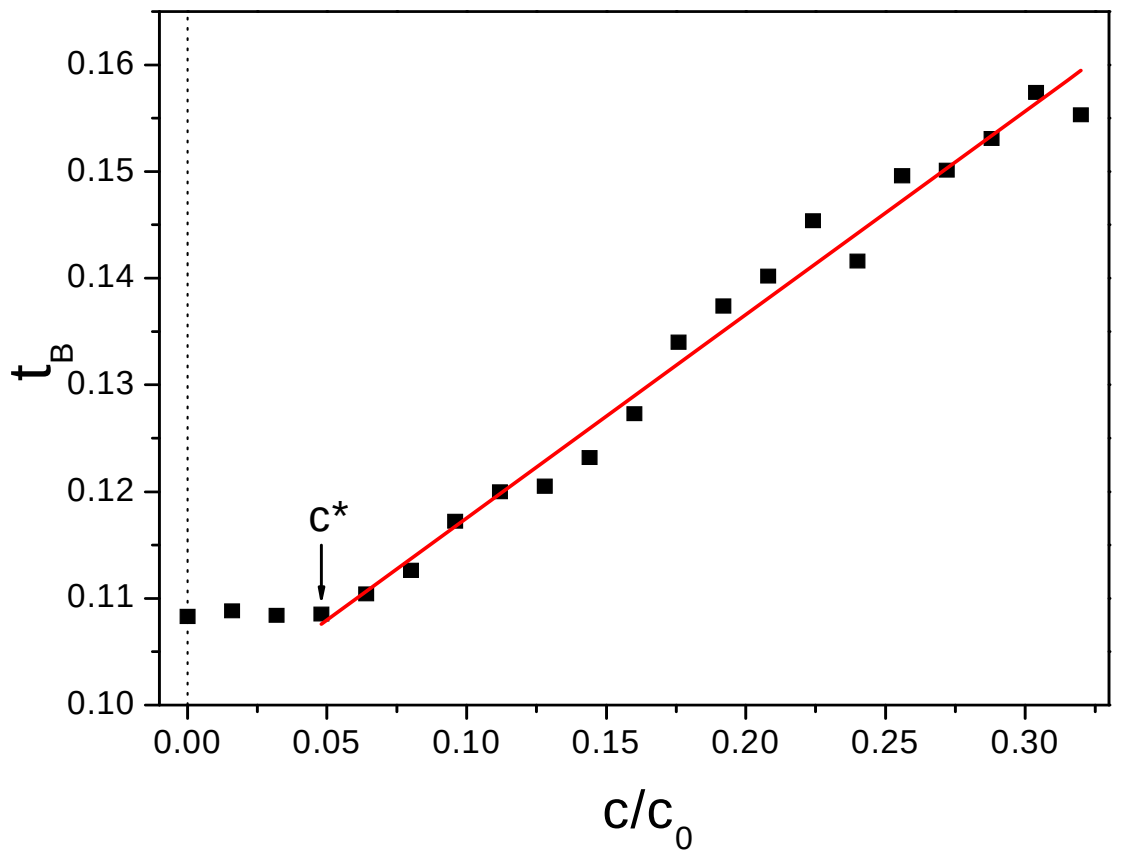

FIG. 2. Plot of $t_B$ as a function of $c/c_0$. The solid line is a fitting to the modified single particle approach (see subsection D). The arrow indicates the concentration c*.

From Fig. 2, we observe a non-monotonic dependence of $t_B$ with concentration: there is a clear change of the slope from an independent $t_B$ at low concentrations to a rapid increase at high concentrations. These different features suggest the presence of two different physical behaviors, an essentially non-interacting regime at low concentrations and an interacting regime at high values of $c/c_0$. The crossover between both regimes is marked by a particular concentration, estimated to be $c^*/c_0 \approx 0.05$. In the next subsections we show some tests we have done to analyze the characteristics of the observed two-regime feature.

**A. Time dependence.**

As the SPM $t_B$ is highly time-dependent,[13] we have simulated the different processes in Fig. 1 at different time intervals (different MC steps), in order to rule out the existence of the two regimes to be a time dependent effect. We varied the MC simulation intervals and maintained the same cooling/heating temperature step. The simulated intervals correspond to 20 and 50 MC steps. The results are shown in Fig. 3, together with the 200 MC steps case of Fig. 2 for the comparison.

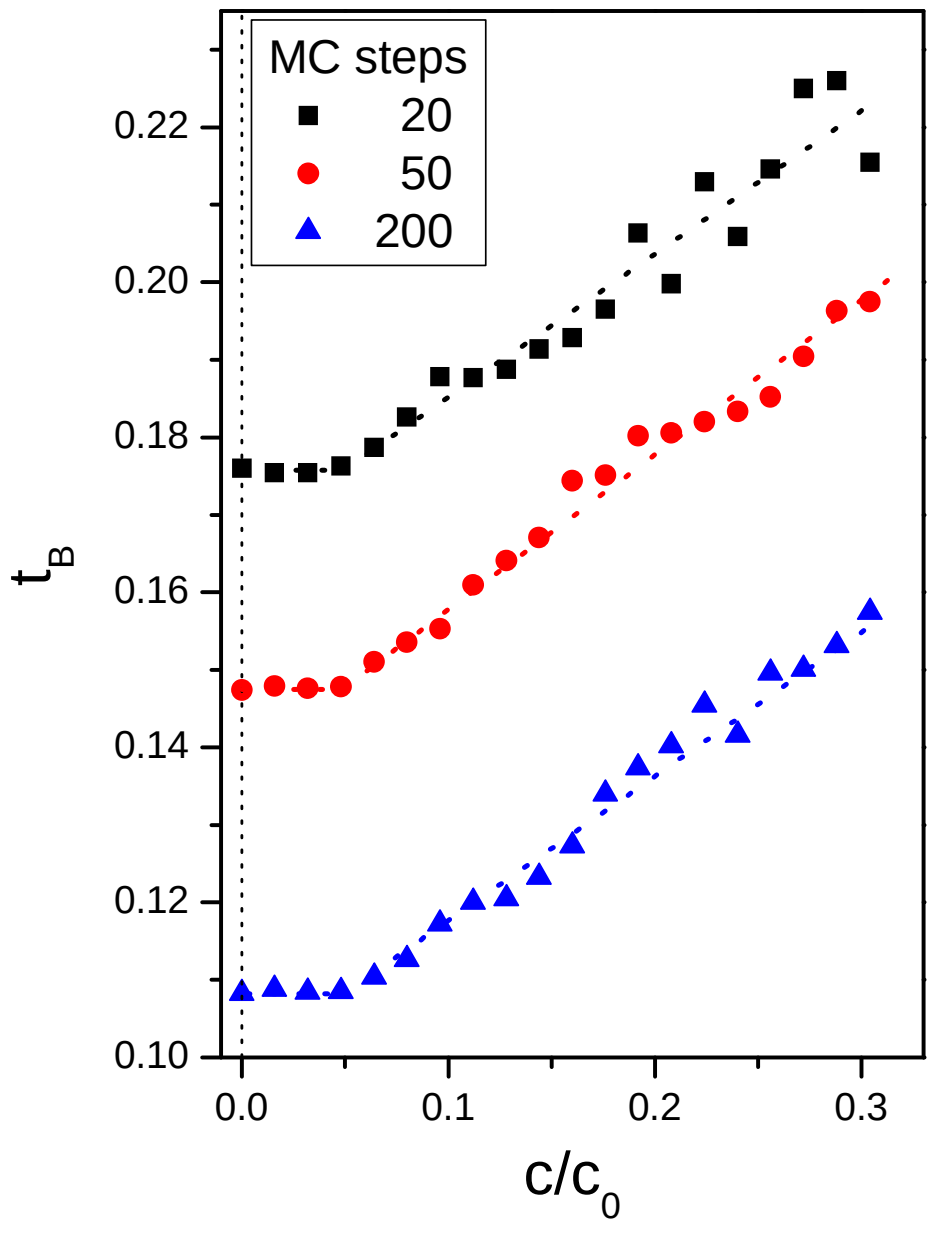

FIG. 3. Plot of $t_B$ vs. $c/c_0$ for different simulation times (MC steps). The dotted lines are guides to the eye.

Seemingly, the overall tendency is the same for the different measuring times: at low concentrations $t_B$ remains basically constant, suggesting the particles to behave independently of each other as a non-interacting system; at higher concentrations $t_B$ increases continuously with the concentration. Hence, we have confirmed that the evolution of $t_B$ with $c/c_0$ is robust for different time intervals.

**B. Maxima at $T_B$.**

To test the existence of the two different regimes we have also analyzed the relative values of the susceptibility at the maximum of the ZFC curves, $\chi(t_B)$, as a function of concentration. This study constitutes a more precise approach than the analysis of $t_B$ because of the higher accuracy found on its determination (see Fig. 1). As the overall observed trend results equal for the different MC step intervals we have focused our study on the 200 MC step case because the simulated ZFC processes show a better definition after more MC relaxation steps. The obtained results are shown in Fig. 4.

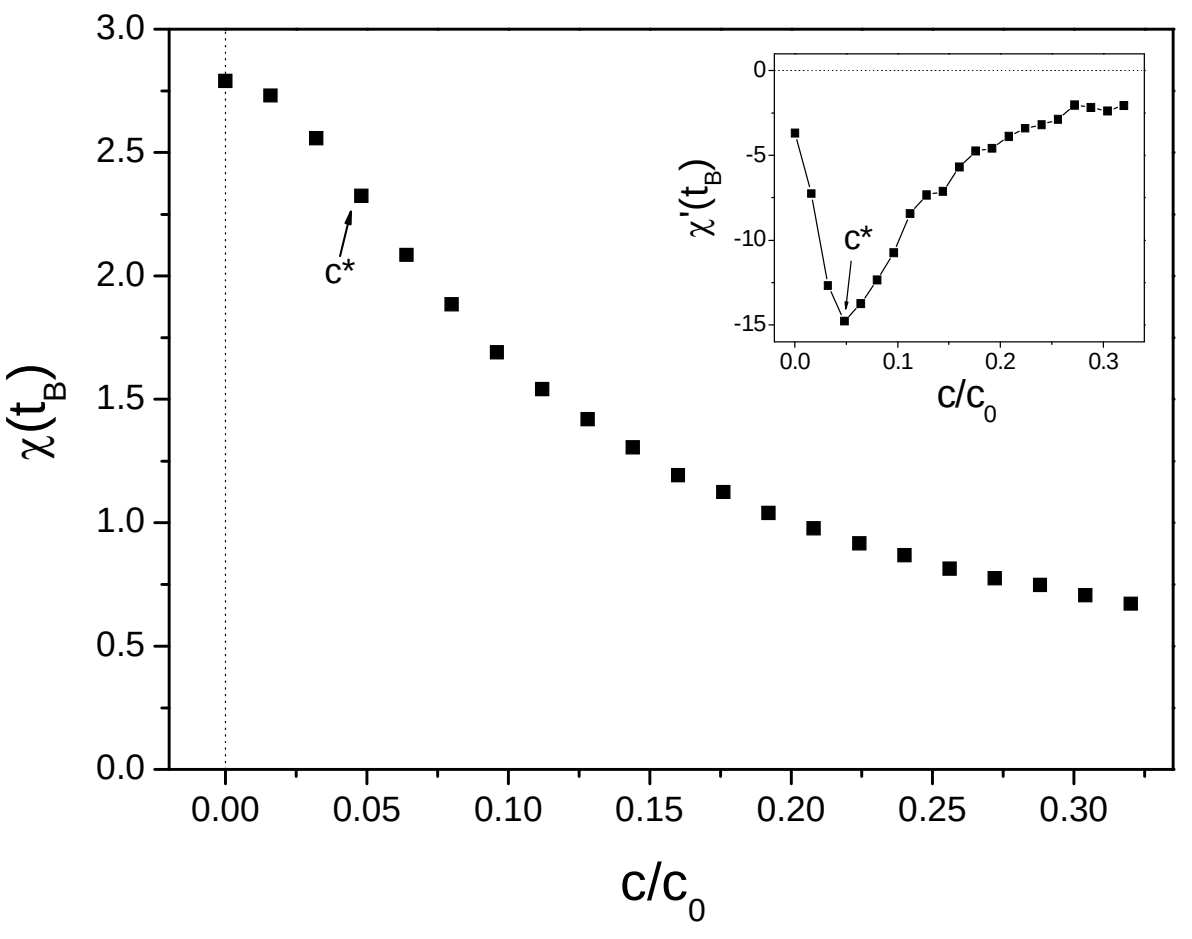

FIG 4. Plot of the reduced susceptibility at the maximum of the ZFC curves for the different interacting conditions considered. Inset shows the first derivative at the maximum of the ZFC curves, $\chi'(t_B)$, as a function of $c/c_0$.

The curve $\chi(t_B)$ vs. $c/c_0$ exhibits an inflexion at low concentrations, as observed in its first derivative (see inset of Fig. 4): an unambiguous minimum appears in $\chi'(t_B)$ at the same reduced sample concentration value $c^*/c_0 \approx 0.05$ observed in Fig. 2. This inflexion on the $\chi(t_B)$ vs. $c/c_0$ curve is related to a change in the magnetic behavior, and supports the existence of two intrinsic regimes of different magnetic behavior in a system of SPM-NPs being influenced by the dipolar interaction.

**C. Comparison with the experiment.**

We report now experimental results extracted from the literature for similar systems but different particle sizes in order to analyze the size-dependence of the trend reported. A similar low-concentration behavior as that shown in Fig. 2 has been observed for highly diluted samples of a frozen ferrofluid of maghemite NPs of ~7 nm diameter.[14] The existence of a defined interparticle spacing separating two different regimes on the evolution of $T_B$ has been reported for iron oxide NPs of ~5.4 nm diameter.[15] With the purpose of probing the generality of the observed behavior, we reproduce in Fig. 5 our results[9] on the evolution of $T_B$ with the sample concentration for a system of magnetic NPs of smaller diameter (~3.5 nm) than those reported in Refs. 14, 15.

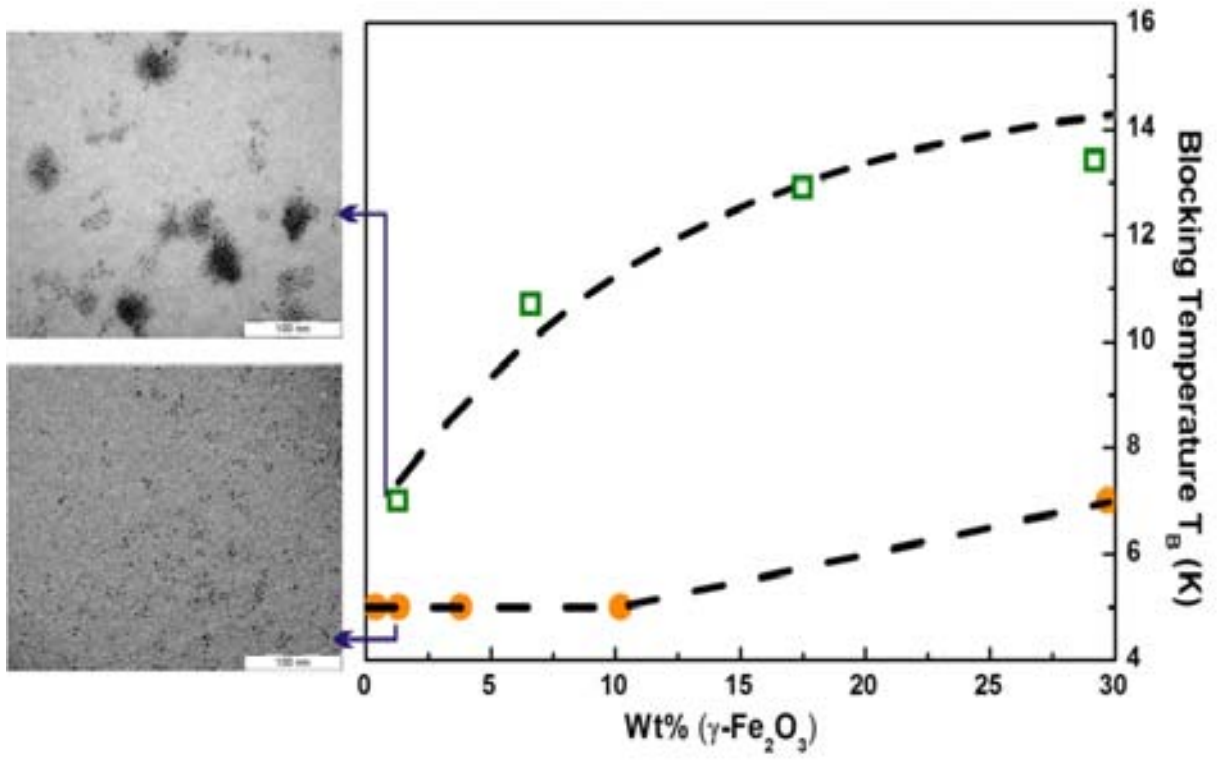

FIG. 5. $T_B$ vs. sample concentration for γ-$Fe_2O_3$ NPs with different aggregation level: a well dispersed sample (open squares) and a highly aggregated one (full circles). The lines serve only to guide the eye. TEM micrographs show the dispersion of the particles for the well dispersed sample (bottom) and for the aggregated one (top) for the same concentration of particles, 1.3 wt% of γ-$Fe_2O_3$.

In Fig. 5, we show two cases of different aggregation level of the same NP system: a well dispersed sample (full circles, left bottom TEM micrograph), and a highly aggregated one (empty squares, left top TEM micrograph) (see Ref. 9 for further details). The presence of a non-interacting regime at low concentrations can be observed in these systems only in the well-dispersed sample (full circles). Presence of clustering (empty squares) clearly affects the shape of the curve and hampers the observation of this regime.[9] These results show that the existence of two different interacting regimes is an intrinsic property of maghemite-like NPs without aggregation, and that is independent of the NP size.

**D. Fitting to classical models.**

Different models have been developed with the purpose to take into account the effect of interparticle interactions on the behavior of the magnetic nanoparticle assemblies. The first attempts were based on modifications of the superparamagnetic single-particle's model by Nèel.[16] In such treatment the interactions between the particles are introduced as changes of the height of the energy barrier, where the different energetic terms add to the anisotropy one. It results in an increment of the thermal activation energy necessary to reach the superparamagnetic state.[17] Recently, W. C. Nunes et al.[11] have proposed a modification of the *Random Anisotropy Model* (RAM) that takes into account the concentration and size of the nanoparticles, as well as the field dependence of the correlation length. We have used both models to fit our results, with the purpose of having one more test to check our arguments. The fitting to both approaches is shown in Fig. 6.

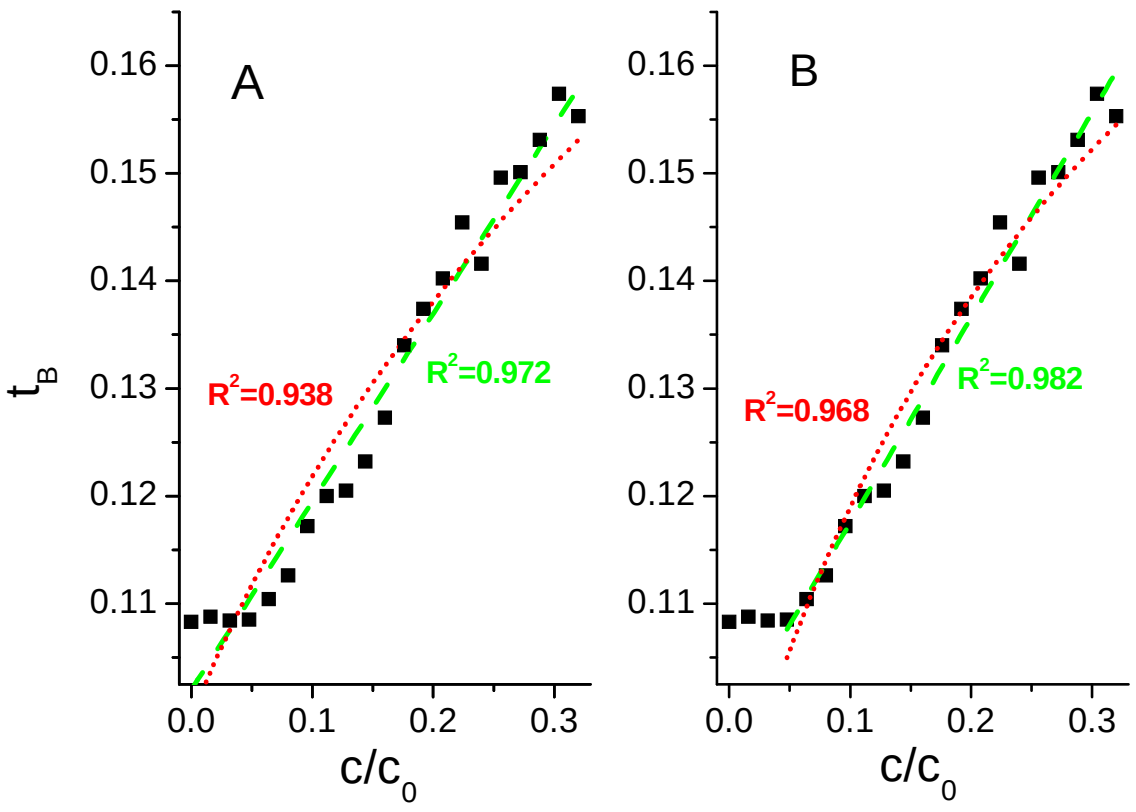

FIG. 6. Fitting of the $t_B$ vs. $c/c_0$ curve to the modified single particle approach (dashed green lines) and the modified RAM approach (red dotted lines). In Fig. 6A the whole range of values is fitted, while in Fig. 6B only the data corresponding to the interacting regime is included in the fitting.

In Fig. 6A the fitting of the whole-range data (the two regimes) is included, while in Fig. 6B only the interacting regime ($c \geq c^*$) data is fitted. The (green) dashed lines correspond to the modified single particle approach, and the (red) dotted lines correspond to the modification of the RAM. The square of the correlation coefficient ($R^2$) is shown in the two fittings for both models. It is clearly observed that the fitting gives very satisfactory results in the interacting range and deviates completely from the expectations at low concentration, since it improves the $R^2$ value for both approaches. This result in fact gives an additional support to our arguments of the existence of two different regimes of behavior of the blocking temperature as a function of the dipolar interaction.

**E. Additional Monte Carlo simulations.**

In order to assure the independency of the reported results on the system size, we have simulated the evolution of $t_B$ with sample concentration using a much larger system, of 1000 particles. Due to computational constraints with this large system size, the temperature variation ratio had to be enlarged to $\Delta t$=0.005000KV/$k_B$ every 500 MC steps and the results were averaged over 300 different configurations. We have concentrated on the values around c*, aiming to focus on the two-regime threshold feature. The values of $t_B$, $\chi(t_B)$, and $\chi'(t_B)$ were evaluated, and the results are shown in Fig. 7.

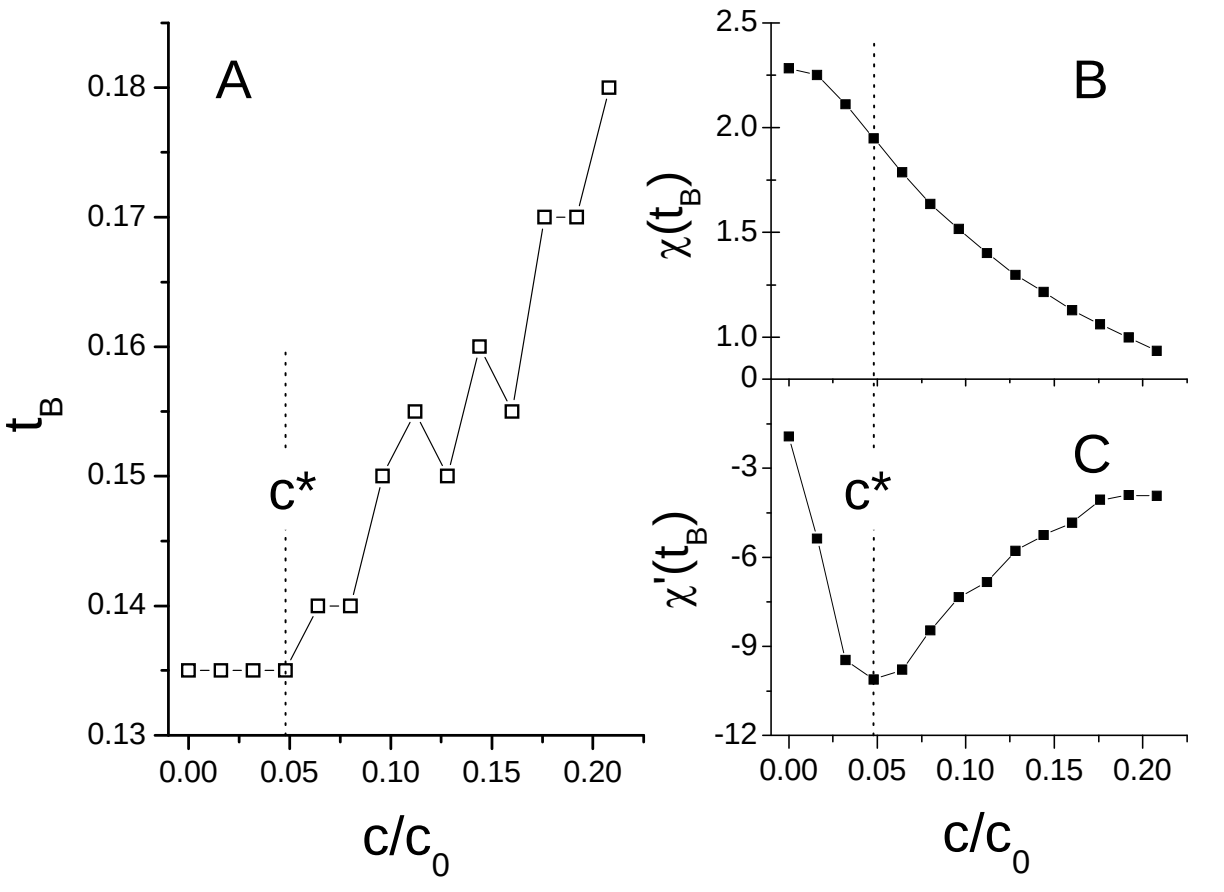

FIG. 7. The evolution of $t_B$ as a function of $c/c_0$ for the N=1000 sample is shown in Fig. 7A, and the corresponding reduced susceptibility and first derivative are plotted in Fig. 7B and Fig. 7C, respectively. Vertical dotted lines indicates the concentration c* extracted from Fig. 2 and Fig. 4.

The results plotted in Fig. 7 indicate that the existence of the two-regime feature discussed in the previous subsections is also observed with the 1000 particles' system and the different temperature interval variation. This demonstrates that the two-regime feature discussed on the previous subsections is independent of the system size used for the simulation. The number of particles considered initially (N=125) is appropriate to study the influence of dipolar interaction on such nanoparticles samples.

**IV. SUMMARY**

Monte Carlo simulations of dispersions of maghemite-like NPs demonstrate a discontinuous evolution of the blocking temperature as a function of the sample concentration that stands for two different interacting regimes. The crossover between these two regimes is determined by a critical concentration c*. At low concentrations ($c < c^*$) $T_B$ remains basically constant (non-interacting regime), while at high concentrations ($c > c^*$) continually increases (interacting regime). This feature has been intensively discussed and analyzed, finding that: i) it is robust for different time intervals; ii) the maxima at $T_B$ also shows an inflexion at c*, what stands for two different behaviors; iii) experimental results show the same tendency and corroborate that its shape is independent on the particle size; iv) our results are in good agreement with the classical interparticle models only in the interacting regime. Moreover, we have also assessed the independency of the reported results on the system size, performing a simulation of the evolution of $t_B$ with sample concentration for a very large system consisting of 1000 particles. On the basis of the results presented here many of the data of the evolution of $T_B$ with concentration in the literature should be revised, considering the possibility of clustering of NPs.

**ACKNOWLEDGEMENTS**

We acknowledge the Xunta de Galicia for Project. No. INCITE 08PXIB236052PR, and for the financial support of D. S. and M. P. (Maria Barbeito and Isabel Barreto programs, respectively). We also acknowledge the Spanish Ministry of Education and Science (Projects No. NAN2004-09203-C04-04, NAN2004-09195-C04-01 and Consolider-Ingenio 2010), and C. E. Hoppe thanks the 2006-IIF Marie Curie Grant (Contract Nº: MIF2-CT-2006-021689, AnaPhaSeS). We thank the CESGA for computational facilities.